\documentstyle[12pt]{article}
\normalsize
\def\simless{\mathbin{\lower 3pt\hbox{$\rlap{\raise 5pt\hbox{$\char'074$}}
\mathchar"7218$}}}
\def\simgreat{\mathbin{\lower 3pt\hbox{$\rlap{\raise 5pt \hbox{$\char'076$}}
\mathchar"7218$}}}
\catcode`@=11

\def\beqra{\begin{eqnarray}} \def\eeqra{\end{eqnarray}}
\def\beq{\begin{equation}}      \def\eeq{\end{equation}}

\def\fo{\hbox{{1}\kern-.25em\hbox{l}}}

\def\ch{\@startsection{section}{1}{\z@}{-3ex plus-1ex minus-.2ex}%
        {2ex plus.2ex}{\large\sc}}

\def\; \lapp \;{\raisebox{-.4ex}{\rlap{$\sim$}} \raisebox{.4ex}{$<$}}

\def\con{\ifmmode \hbox{\bf*} \else{\bf*}\fi}   
\def\scon{\ifmmode \hbox{\footnotesize\rm\bf*} \else{\footnotesize\rm\bf*}\fi}

\def\0#1{\relax\ifmmode\mathaccent"7017{#1}
        \else\accent23#1\relax\fi}              

\def\eslash{\not{\hbox{\kern-2pt $E$}}}

\catcode`@=12

\begin{document}
\hoffset=0.4cm
\voffset=-1truecm
\normalsize


\begin{titlepage}
\begin{flushright}
UMD-PP-94-102
\end{flushright}
\begin{flushright}
March,1994
\end{flushright}
\vspace{24pt}
\centerline{\Large {\bf A Single Scale Theory for Cold and Hot Dark Matter }}
\vspace{24pt}
\begin{center}
{\large\bf R.N.
Mohapatra$^{a}$\footnote{Work supported by a grant from the National
Science Foundation; Email: rmohapatra@umdhep.umd.edu}
 and A. Riotto$^{b,c}$\footnote{Email:riotto@tsmi19.sissa.it}}
\end{center}
\vskip 0.2 cm

\centerline{\it $^{(a)}$Department of Physics and Astronomy}
\centerline{\it University of Maryland, College Park, MD 20742}
\vskip 0.2 cm
\centerline{\it $^{(b)}$International School for Advanced Studies,
SISSA-ISAS}
\centerline{\it Strada Costiera 11, 34014 Miramare, Trieste, Italy}
\vskip 0.2 cm
\centerline{\it $^{(c)}$Istituto Nazionale di Fisica Nucleare,}
\centerline{\it Sezione di Padova, 35100 Padua, Italy.}
\vskip 0.2 cm

\vskip 0.5 cm
\centerline{\large\bf Abstract}
\vskip 0.2 cm
\baselineskip=24pt
   We show that a recently proposed extension of the minimal
supersymmetric standard model
 can provide a  scenario where both the cold and the hot
 dark matter of the universe owe their origin to a single scale
connected with the breakdown of global $B-L$ symmetry. The
susy partner of the majoron and the light Majorana neutrinos
are the cold and hot dark matter candidates respectively
in this model and
their desired relative abundances
emerge when the scale of $B-L$ symmetry breaking is in the TeV
range.

\end{titlepage}
\baselineskip=24pt

The idea that the large scale structures seen today evolved from very
small primordial density inhomogeneities has been strengthened by the
recent detection of large scale anisotropies in the cosmic microwave
background \cite{smoot}. Nevertheless, one of the necessary ingredients
for the structure formation, namely the nature of dark matter, remains
unknown.

The most satisfactory model for structure formation is perhaps the cold
dark matter (CDM) theory \cite{davis} where the Universe is assumed to
be spatially flat ($\Omega=1$) and with $\sim 0.9$ of the mass density
formed by CDM particles, cold in the sense that they decouple from the
expanding thermal bath at temperatures much smaller than their mass. CDM
can successfully explains galaxy-galaxy and cluster-cluster correlation
functions on scales of order of 1-5 Mpc. However, it now appears to be
inconsistent with large scale structure data in the automatic plate
machine (APM) galaxy survey \cite{apm}, which suggest more power on large
scales than the standard CDM predictions. On small scales, the observed
pairwise velocity dispersion for galaxies appears to be smaller than
those predicted by CDM \cite{pee}.

One alternative is the hot dark matter model (HDM). HDM is taken to be a
light neutrino, which decouples from the thermal bath when still
relativistic, with $m_\nu=(92\:\Omega_\nu\:h^2)$ eV where $H=100\: h$
Km/sec Mpc is the Hubble parameter. In the pure HDM picture,
 the processed fluctuation is
characterized by the distance a neutrino travels over the history of the
Universe, $\lambda_\nu\simeq 40$ (30 eV/$m_\nu$) Mpc. The problem with
this however is that $\lambda_\nu$ is too large with respect to the scale
which is just becoming nonlinear today, $\sim 5\: h^{-1}$ Mpc. If galaxy
formation occurs early enough to be consistent with high-redshift
galaxies and quasars, then structures on $ 5\: h^{-1}$ Mpc will
overdevelop.

The hope is that cold + hot dark matter (C + HDM) will combine the
success of both models. Indeed (C + HDM) models with $\Omega_{CDM}\simeq
0.6$, $\Omega_{\nu}\simeq 0.3$, $\Omega_{baryon}\simeq 0.1$ and a
Hubble constant $h\simeq 0.5$ provide the best fit for microwave anisotropy
data, large scale structure surveys, and measures of the bulk flow with
a few hundred megaparsecs \cite{mixed}.

Even if C + HDM is appealing for the large scale structure
phenomenology, it
might seem rather unpalatable from the point of view of
particle physics for the following reasons. It is well-known that
low-energy supersymmetric theories provide an elegant solution to the
hierarchy problem \cite{susy}. The lightest supersymmetric particle
(LSP) is the most attractive candidate for CDM and is made stable by
imposing a discrete symmetry, called $R$-parity. Nevertheless, in the
minimal supersymmetric standard model (MSSM), neutrinos are not massive
and therefore there are no candidates for HDM. If one considers, for
instance, minimal extension of the MSSM by including a right-handed neutrino
superfield $\nu^c$ and a singlet field $S$ with lepton number $L(S)=+2$
\cite{giudice}, the $U(1)_{B-L}$ and $R$-parity symmetries get spontaneously
broken, neutrinos acquire a mass, but the LSP is no longer stable and
cannot play the role of CDM. So, when trying to build up a
supersymmetric model with both CDM and HDM components, one has to face
the problem of making neutrinos massive
 while at the same time, keeping the LSP
stable. Moreover, even if one is successful in doing so, the relative
abundances of light massive neutrinos and LSP's are expected to be set
by two uncorrelated scales, namely the $U(1)_{B-L}$ breaking scale,
$V_{BL}$, and the supersymmetric breaking scale, $M_S$, and a sort of
fine-tuning on $V_{BL}$ and $M_S$ must be done to satisfy the
requirements of C + HDM model.

In this Letter, we give an example of an extension of the MSSM, where it is
possible to implement  C + HDM model for large scale structure formation
without arbitraily separating the scales  $V_{BL}$ and $M_S$.
The cold dark matter in this model is the lightest supersymmetric
particle of the theory ( not the usual neutralino ) with mass in the
range of 10 to 50 GeV while the hot dark matter is an eV Majorana neutrino.
What we find quite interesting is that
 the relative abundances of the HDM and CDM components are set by
the {\it same} scale $V_{BL}\simeq 10^3$ GeV. The new scale being so low
brings the hope that the effects of this low scale may be testable
in near future.

The model is a supersymmetric version of the
original singlet Majoron model \cite{cmp} and was recently proposed
in ref. \cite{zhang}. For
$V_{BL}\geq M_S$ (where $M_S$ is the scale of supersymmetry (SUSY)
breaking), the susy partners of the majoron (the smajoron and
majorino) acquire mass of $M_S$ . Since they are very weakly coupled to
light particles of the theory such as the quarks and the leptons,
 for their cosmological
relic  abundance not to disturb nucleosynthesis, the scale of
$B-L$ symmetry breaking $V_{BL}$ must have an upper bound of order
of a few TeV\cite{zhang}. In this letter,
 we show that if the majorino is the
lightest superparticle of the model (i.e. lighter than the familiar
neutralino), then due to $R$-parity conservation the majorino is stable
and for $V_{BL}\simeq$ TeV, has the right relic abundance to be the cold
dark matter. The neutrinos in this model are of course massive with either
the $\nu_e$ or $\nu_\mu$ or both
 having  mass in the eV range and  therefore becoming the
obvious hot dark matter candidate. As far as the  $\nu_{\tau}$
is concerned, cosmological constraints demand that in our model
its mass either be in the eV range ( thereby sharing the dark-matter-dom
with other neutrinos ) or in the MeV range in which case it decays
to $\nu_{e,\mu}$~+~Majoron fast enough so as not to effect
structure formation in the universe.

Let us now proceed to discuss the details of the model.
The superpotential can be written as the sum of two terms
\beq
W=W_0 + W_1,
\eeq
where $W_0$ is the usual MSSM piece and (generation indices are
suppressed)
\beq
W_1= h_\nu L H_2 \nu^c + f \nu^c \nu^c S_1+
\lambda\left(S_1S_2-M^2\right)Z.
\eeq
Here $M$ is an explicit mass scale of order of $V_{BL}$;
$\nu^c$ is a right-handed neutrino superfield which is a singlet
under $SU(2)_L\otimes U(1)_Y$ and carries lepton number $L=-1$;
$S_1$, $S_2$ and $Z$ are $SU(2)_L\otimes U(1)_Y$ singlet superfields as
well and carry lepton numbers $L=2$, $-2$ and $0$, respectively.

It easy to work out from eqs. (1) and (2) the full scalar potential
\cite{susy}. In particular, we are interested in the pieces containing
the fields $\nu^c$, $S_1$, $S_2$ and $Z$ (we are assuming $h_\nu\ll
\lambda,f$ and all the couplings real for simplicity)
\begin{eqnarray}
V&=& 4 f^2\left|\tilde{\nu}^c\right|^2\left|S_1\right|^2+\lambda^2
\left|S_1S_2-M^2\right|^2+\lambda^2\left|S_1\right|^2\left|Z\right|^2
\nonumber\\
&+&\left|f\tilde{\nu}^c\tilde{\nu}^c+\lambda S_2 Z\right|^2 +
m_{1}^{2}\left|S_1\right|^2  +
m_{2}^{2}\left|S_2\right|^2 +
m_{Z}^{2}\left|Z\right|^2+
m_{\tilde{\nu}^c}^{2}\left|\tilde{\nu}^c\right|^2\nonumber\\
&+&\left(A_f \tilde{\nu}^c\tilde{\nu}
^c S_1+{\rm h.c.}\right)+\left(A_{\lambda} S_1S_2
Z+ {\rm h.c.}\right)-\left(B_{\lambda} M^2 Z+{\rm h.c.}\right).
\end{eqnarray}
In eq. (3)
$\left(m_1,m_2,m_Z,m_{\nu^c},A_f,A_{\lambda},B_{\lambda}\right)$ are all
parameters which set the scale of the soft breaking of supersymmetry and
we assume them to be of the same order of magnitude, $M_S\simeq
10^2$ GeV.

One would naively expect that for temperatures below $M_S$,
the symmetry breaking effects would be negligible in analogy with what
happens in spontaneously broken gauge symmetries such as electroweak.
However, the above symmetry breaking is {\it not} spontaneous but
explicit and there is no reason to expect the restoration of the
symmetry above $M_S$. Rather, the relevant question is above which
temperature, the effect of the soft symmetry breaking terms falls out of
equilibrium. One could estimate easily this temperature by comparing the
rate for the symmetry breaking effects $\Gamma_S\simeq M_{S}^{2}/T$ with
the expansion rate of the Universe, $H\simeq g_{*}(T)^{1/2}T^2/M_{Pl}$,
where $g_{*}(T)$ counts the effective number of relativistic degrees of
freedom and $M_{Pl}\simeq 1.2\times 10^{19}$ GeV is the Planck mass.
One finds that the symmetry breaking effects are out of equilibrium for
\beq
T\simgreat T_S\simeq
\frac{M_{S}^{2/3} M_{Pl}^{1/3}}{30^{1/3}}\simeq 10^7 \:\mbox{GeV}.
\eeq
Since we are interested in values of $V_{BL}$ not far from $M_S$ (see
later) and certainly in the range $M_S\simless V_{BL}\simless
T_S$, we cannot neglect in the scalar potential (3) the soft
supersymmetric breaking terms.

As the temperature falls  below the value $T\simeq V_{BL}$, the
$U(1)_{B-L}$ symmetry is spontaneously broken by the following vacuum
expectation values (VEV's)
\beq
\langle S_1\rangle=V_1, \:\:\:\:\:\:\langle S_2\rangle=V_2, \:\:\:\:\:\:
V_1\simeq V_2\simeq V_{BL},
\eeq
whereas it is easy to show that $Z$ acquires a VEV
\beq
\langle Z\rangle =  \frac{B_{\lambda} M^2-A_{\lambda}V_1
V_2}{\lambda^2\left(V_{1}^{2}+V_{2}^{2}\right)}.
\eeq
The crucial point here is that, whenever $V_{BL}\simgreat M_S$, the
right-handed sneutrino $\tilde{\nu}^c$ does not acquire a VEV so that
the discrete $R$-parity symmetry, which is proportional to
$(-1)^{L}$, is preserved. As a consequence, the LSP in our model remains
stable and can provide a suitable candidate for CDM.

After the spontaneous breaking of the $U(1)_{B-L}$ symmetry, a
Nambu-Goldstone boson, the majoron, will appear
\beq
J=\frac{V_1\left({\rm Im}\:S_1\right)-V_2\left({\rm Im}\: S_2\right)}{
\sqrt{V_{1}^2+V_{2}^2}},
\eeq
whereas its fermionic superpartner, the majorino
\beq
\psi_J=\frac{V_1 \psi_1-V_2\psi_2}{
\sqrt{V_{1}^2+V_{2}^2}}
\eeq
and its real superpartener, the smajoron
\beq
\sigma_J=\frac{V_1\left({\rm Re}\:S_1\right)-V_2\left({\rm Re} \:S_2\right)}{
\sqrt{V_{1}^2+V_{2}^2}}
\eeq
will acquire a mass proportional to $M_S$. In particular, the majorino
$\psi_J$ receives two different mass contributions: one, at the tree
level, of order of $\lambda\langle Z\rangle$, and the second from
one-loop diagrams involving $\tilde{\nu}^c$ and $\nu^c$ of order of $(f
M_S/16 \pi^2)$. Without any fine-tuning of the parameters, we can have
$m_{\psi_J}\simeq \left(10-50\right)$ GeV and consider the majorino
lighter than any other neutralino. As a consequence, the majorino will
be  the LSP and a suitable candidate for CDM.
The smajoron being heavier can decay through various channels as discussed
in ref. \cite{zhang}. We have  found a new decay channel $\sigma_J\rightarrow
JJ$ \cite{chun}, which provides a somewhat faster decay mode. We find that,
this decay amplitude goes like $\simeq \left(M^{2}_{S}/ V^{3}_{BL}\right)$
and requiring the $\sigma_J$ lifetime to be $\simeq 10^{-2}$ sec. in order
not to disturb neucleosynthesis discussion implies that $V_{BL}\leq 10^6$ GeV.
Our discussion of majorino as the CDM needs $V_{BL}\simeq$ TeV, which is
consistent with the above bound.

We now proceed to the estimation of the relic abundance of majorinos.
We first want  to point out that the couplings of the majorino,
 the smajoron and
the majoron among themselves and with the other fields must be
calculated in presence of the soft supersymmetry breaking terms and not
in the exact limit of supersymmetry ,since
we are considering temperatures below $T_S$.
For $T\simless V_{BL}$ all the heavy fields, such as $\tilde{\nu^c}$,  decay
into lighter particles so that the majorino, as well as the smajoron and
the majoron, can only interact among each other and with the "standard
model" particles, such as leptons and quarks, through the coupling
$h_\nu L H_1 \nu^c$. Therefore, the coupling of $\psi_J's$ and $J's$
to the "standard model" particles is suppressed by powers of
$\epsilon\simeq(f^2 h_{\nu}^2/16\pi^2)$ and they decouple from the
"standard model" thermal bath at a temperature \cite{zhang}
\beq
T_{\gamma D}\simeq 10^4 \:V_{BL}\left(\frac{V_{BL}}{M_{Pl}}\right)^{1/3}
\left(\frac{10^{-6} f^2}{\epsilon}\right)^{2/3}\: \mbox{GeV}.
\eeq
To get a feeling of the numbers, $T_{\gamma D}\simeq 10^2$ GeV for
$V_{BL}\simeq 10^3$ GeV. Therefore, the majorino is expected to decouple
from the "standard model" thermal bath when still relativistic.
Nevertheless, its number density for $T\simless T_{\gamma D}$ does not
decrease only due to the expansion of the Universe. Indeed, the key
point here is that, even after $T_{\gamma D}$, the
 number density of majorinos continue to follow its
equilibrium value due to the fact that they keep into equilibrium with the
thermal bath formed by majorons via interactions mediated by heavy
particles with mass $\sim V_{BL}$ such as Re$Z$ or the two fermionic
combinations of $\psi_1$, $\psi_2$ and $\psi_Z$ orthogonal to
$\psi_J$.

If we denote by $T_J$ the temperature of the thermal bath formed
by majorons and majorinos and by $T_{\gamma}$ the one relative to the
"standard model" particles, for $T_{\gamma}<T_{\gamma D}$ we do have
$T_J=a(T_{\gamma})T_{\gamma}$ where
\beq
a\left(T_{\gamma}\right)\equiv\left(\frac{g_{*S}(T_{\gamma})}{
g_{*S}(T_{\gamma D})}\right)^{1/3}
\eeq
takes into account the various annihilation thresholds for massive
"standard model" particles and $g_{*S}(T_{\gamma})$ counts the effective
relativistic degrees of freedom contributing to the entropy density.

It is  easy to show that majorinos can annihilate into a pair
of majorons only through a $p$-wave and denoting the thermally averaged
cross section for the process $\psi_J\psi_J\rightarrow JJ$ by
$\langle\sigma|v|\rangle=\sigma_{0}\times(T_{J}/m_{\psi_{J}})$, we can
calculate the temperature $T^{*}_{\gamma}$
at which majorinos freeze out from the thermal bath of majorons by
simply comparing the interaction rate
\beq
\Gamma_{int}= n_{\psi_{J}}^{EQ}
\langle\sigma|v|\rangle=2\left(\frac{m_{\psi_J}T_{J}}{2\pi}\right)
^{3/2}{\rm e}^{
-m_{\psi_J}/T_{J}}\sigma_0\times\left({{T_J}\over{m_{\psi_J}}}\right)
\eeq
 with the expansion rate of the Universe. Here we do
not make use of the Boltzmann equation for $n_{\psi_J}$ since it would
give a result different from ours only for a few percent. The freeze
out temperature is then
\beq
T_{\gamma}^{*}\simeq\frac{m_{\psi_J}}{a(T_{\gamma}^{*})}
\left[{\rm ln} A-\frac{1}{2}{\rm ln}\left({\rm ln}A\right)\right]^{-1},
\eeq
where
\beq
A\simeq\left[\frac{\sigma_{0}[a\left(T_{\gamma}^{*}\right)]^{5/2}}
{g_{*}\left(T_{\gamma}^{*}\right)^{1/2}} 0.038 \:m_{\psi_{J}} M_{Pl}\right].
\eeq
The today contribution of majorinos to the $\Omega$ parameter
is then given by \cite{comment1}
\begin{eqnarray}
\Omega_{\psi_{J}}h^2&\simeq& 2.4\:\times 10^{46}\:
m_{\psi_{J}}\left(\frac{m_{\psi_J}}{2\pi T_{\gamma}^{*}}\right)^{3/2}
{\rm
e}^{-m_{\psi_J}/a(T_{\gamma}^{*})T_{\gamma}^{*}}\nonumber\\
&\times& a(T_{\gamma}^{*})^{3/2}\:
T_{\gamma,tod}^3 \:\frac{g_{*S}(T_{\gamma,tod})}{g_{*S}(T_{\gamma}^{*})}
\:\mbox{GeV}^{-4},
\end{eqnarray}
where $T_{\gamma,tod}\simeq 2.75$ K is the today temperature of the
relic photons.

Taking $h\simeq 0.5$ and $\Omega_{\psi_{J}}\simeq 0.6$ as suggested by
the latest C + HDM simulations and $a(T^{*}_\gamma)\simeq .5$,
 we get $\sigma_0\simeq 10^{-11}$
GeV$^{-2}$. Since a detailed calculation of $\sigma_0$ with standard
techniques \cite{griest} gives, for $\lambda\simeq
0.5$,  $\sigma_{0}\simeq 2\times 10^{-3}
\: (m_{\psi_J}^{2}/V_{BL}^{4})$ \cite{comment2},
we obtain that majorinos can form the CDM
component of the Universe for a relativley small value of $V_{BL}$,
$V_{BL}\simeq 10^{3}$ GeV. Such a small value of $V_{BL}$ is quite
likely to manifest itself in rare decay processes. However,
in the model as presented, it would be
rather difficult to detect the CDM component either through direct
searches or through indirect detection of annihilation products
of majorinos that annihilate in the Sun, in the Earth or in
the galactic halo since the majorino is
very weakly coupled to matter. It is however possible to add to
the superpotential in Eq. (2) terms like $ZH_1H_d~ +~Z^3 $, which then allow
the Majorino to have somewhat stronger interaction with matter,
while still remaining a viable CDM. This will be discussed in more detail
in a forthcoming publication \cite{mr2}.

 Let us now turn to the nature of the neutrino spectrum. As it is well-known,
the neutrino mass matrix in this model has the familiar see-saw form
and leads to neutrino masses given by $m_{\nu_i}\simeq
\left(h^2_{\nu_i}v^2_{WK}/
V_{BL}\right)$. The nature of the light neutrino spectrum depends
crucially on the choice of $h_{\nu_i}$'s, which are apriori arbitrary.
If we choose the Yukawa couplings $h_{\nu_i}\simeq 10^{-5}$ (which is
comparable to the Yukawa coupling of the electron and is therefore
not unreasonable), then all three
neutrino species have masses in the eV range and get to
 play the role of the hot dark matter together. On the other hand, it
is perhaps more suggestive to have $h_{\nu_i}v_{WK}\simeq m_{l_i}$,
in which case, one may get $m_{\nu_\mu}\simeq 100$ to 200 keV and
$m_{\nu_\tau}\simeq$ 5 to 10 MeV. In our model, the heavier neutrinos
(i.e. $\nu_\mu$ and $\nu_\tau$) decay to the lighter neutrino plus a
Majoron. Their decay is suppressed in the lowest order \cite{val} and
arises in order $\approx (m_{\nu_i}/V_{BL})^2 $. These decay rates
have been calculated in detail in ref. \cite{cline}, where
it is found that for $\nu_i\rightarrow \nu_j~+~ J$ decay, the lifetime
$\tau_{ij}$ is given by
\beq
       \tau^{-1}_{ij}\simeq {\rm sin}^2{2\alpha}
 {{m_{\nu_i} m^4_{\nu_j}}\over{16\pi~V^4_{BL}}}
\eeq

Here, $\alpha$ is a rotation angle in the Majoron coupling matrix and
being unrelated to the observed neutrino mixing angles, its value can
be of order one, although we will allow it to be anywhere from 0.1 to 1
in our discussion below.
 From the above equation we see that, for a 10 MeV tau neutrino, both the
cosmological mass density \cite{cowsik}
 as well the galaxy formation \cite{steig} constraints
are easily satisfied and the decay neutrinos , which get redshifted
sufficiently to leave the universe matter-dominated before the
epoch of galaxy formation,
 join the already present relic HDM neutrinos.
Coming to the $\nu_\mu$ since its mass has to be less than
220 keV , it cannot satisfy
either of these constraints. As a result, in our model $\nu_\mu$ must
have a mass in the few eV range contributing to the HDM content.
 Thus, to summarize, the HDM can
contain both the $\nu_e$ and $\nu_{\mu}$ with masses in the 2 to 3 eV
range each whereas the tau neutrino can either be very light and in the
eV range or in the heavy range of several MeV's. In the latter case of
course, it decays to both the electron and muon neutrinos. Also it is
worth noting that, the existence of
a Majorana $\nu_e$ with mass in the 2-3 eV range will be testable in
the ongoing $\beta\beta_{0\nu}$ experiments. Failure to observe such a
mass will imply that in our picture, a few eV muon neutrino becomes the
dominant HDM.

In conclusion, we have given an example of an extension
of the MSSM, where it is
possible to implement the C + HDM scenario for large scale structure
formation using a single new scale for physics beyond the standard
model, i.e.the scale of
 $U(1)_{B-L}$ breaking scale needed anyway to understand the
possible small neutrino masses. It is interesting that, the
relative abundances of the CDM and HDM components are set by the
same scale being in the TeV range, raising the hope that such
theories may be testable in near future.

\vspace{1. cm}

\centerline{\bf Acknowledgments}

It is a pleasure to express our gratitude to A. Masiero
for useful discussions.
\vskip 2.cm
%
\def\MPL #1 #2 #3 {Mod.~Phys.~Lett.~{\bf#1}\ (#3) #2}
\def\NPB #1 #2 #3 {Nucl.~Phys.~{\bf#1}\ (#3) #2}
\def\PLB #1 #2 #3 {Phys.~Lett.~{\bf#1}\ (#3) #2}
\def\PR #1 #2 #3 {Phys.~Rep.~{\bf#1}\ (#3) #2}
\def\PRD #1 #2 #3 {Phys.~Rev.~{\bf#1}\ (#3) #2}
\def\PRL #1 #2 #3 {Phys.~Rev.~Lett.~{\bf#1}\ (#3) #2}
\def\RMP #1 #2 #3 {Rev.~Mod.~Phys.~{\bf#1}\ (#3) #2}
\def\ZP #1 #2 #3 {Z.~Phys.~{\bf#1}\ (#3) #2}

\newpage



\begin{thebibliography}{99}
\bibitem{comment1} We note here that a further contribution to
$\Omega_{\psi_J}$ can come from the next-to-lightest
neutralino $\tilde{\chi}_{1}$, which can produce majorinos through
the decay mode $\tilde{\chi}_{1}\rightarrow \psi_{J}\nu\nu$. The decay
rate of such a process will be proportional to $|\beta|^2$, where
$\beta$ represents the amount of the higgsino $\tilde{H}_2$ in
$\tilde{\chi}_1$. Naturally, this further contribution can be neglected
if the relic abundance of $\tilde{\chi}_1$'s is not significant as
expected for supersymmetric parameters in the 100 GeV range.
\bibitem{comment2} The dominant contribution to $\psi_J\psi_J\rightarrow
JJ$  comes from the fermion intermediate state $\psi_Z$ and its
amplitude goes like $\simeq V^{-2}_{BL}$.
\end{thebibliography}
\end{document}